\documentclass[11pt,a4paper]{article}
\usepackage{jinstpub}

\usepackage[utf8]{inputenc}
\usepackage{physics}
\usepackage{xspace}
\usepackage[utf8]{inputenc}
\usepackage{graphicx}
\usepackage{placeins}
\usepackage{lineno}
\usepackage{units}
\usepackage{multirow}
\usepackage{amsmath}
\usepackage{booktabs}
\usepackage{caption}
\usepackage{subcaption}
\usepackage[notlof,notlot,nottoc]{tocbibind}

\newcommand{\wrap}[1]{\ensuremath{#1}\xspace}
\newcommand{\dt}{\wrap{{\Delta}t}}

\newcommand{\um}{\wrap{\muup\textrm{m}}}
\newcommand{\oC}{\wrap{^{\circ}\textrm{C}}}

\DeclareMathOperator\erfc{erfc}

\title{Threshold bounce --- occupancy-dependent modulation of the discriminating threshold in silicon detectors}

\author[a,b,1,2]{M.J.~Basso,}\emailAdd{mbasso@triumf.ca}
\author[c,3]{E.~Buchanan,}
\author[d]{B.J.~Gallop,}
\author[c,e]{J.J.~John,}
\author[c]{J.~Kaplon,}
\author[f]{P.T.~Keener,}
\author[d]{P.W.~Phillips,}
\author[a,b]{L.~Poley,}
\author[d]{C.A.~Sawyer,}
\author[g]{D.~Sperlich,}
\author[h]{and M.~Warren}

\note{Corresponding author.}
\note{Previously at the University of Toronto.}
\note{Previously at the Chinese Academy of Sciences.}

\affiliation[a]{TRIUMF, Wesbrook Mall, Vancouver, Canada}
\affiliation[b]{Department of Physics, Simon Fraser University, University Drive W, Burnaby, Canada}
\affiliation[c]{Experimental Physics Department, CERN, Geneva, Switzerland}
\affiliation[d]{Particle Physics Department, STFC Rutherford Appleton Laboratory, Harwell Science and Innovation Campus, Didcot, United Kingdom}
\affiliation[e]{Department of Physics, Oxford University, Oxford, United Kingdom}
\affiliation[f]{Department of Physics and Astronomy, University of Pennsylvania, South 33rd Street, Philadelphia, USA}
\affiliation[g]{Physikalisches Institut, Albert-Ludwigs-Universit{\"a}t Freiburg, Hermann-Herder-Stra{\ss}e, Freiburg im Breisgau, Germany}
\affiliation[h]{Department of Physics and Astronomy, University College London, Gower Street, London, United Kingdom}

\abstract{

The front-end electronics of silicon detectors are typically designed to ensure optimal noise performance for the expected input charge. A combination of preamplifiers and shaper circuits result in a nontrivial response of the front-end to injected charge, and the magnitude of the response may be sizeable in readout windows subsequent to that in which the charge was initially injected. The modulation of the discriminator threshold due to the superposition of the front-end response across multiple readout windows is coined ``threshold bounce''.

In this paper, we report a measurement of threshold bounce using silicon modules built for the Phase-II Upgrade of the ATLAS detector at the Large Hadron Collider. These modules utilize ATLAS Binary Chips for their hit readout. The measurement was performed using a micro-focused \unit[15]{keV} photon beam at the Diamond Light Source synchrotron. The effect of the choice of photon flux and discriminator threshold on the magnitude of the threshold bounce is studied. A Monte Carlo simulation which accounts for the front-end behaviour of the silicon modules is developed, and its predicted hit efficiency is found to be in good agreement with the measured hit efficiency.

}

\keywords{Si microstrip and pad detectors; Front-end electronics for detector readout; Radiation-hard electronics}

\begin{document}

\maketitle
\flushbottom

\section{Introduction}
\label{sec:intro}

The High-Luminosity (HL) Upgrade~\cite{ZurbanoFernandez:2020cco} of the Large Hadron Collider (LHC)~\cite{Evans:2008zzb} will result in ten  times more data over its lifetime than its existing design, allowing physicists to probe the fundamental symmetries of the universe with higher precision. To cope with the increased occupancy and radiation dose accompanying the increased flux of proton-proton collisions, the ATLAS detector~\cite{ATLAS:2008xda} is upgrading its Inner Detector~\cite{ATLAS:1997ag, ATLAS:1997af, ATLASIBL:2018gqd, ATLAS:2023dns} as part of its Phase-II Upgrade~\cite{ATLASCollaboration:2012ilu}. The new Inner Detector --- known as the Inner Tracker (ITk) --- will be all-silicon and consist of an inner pixel detector~\cite{ATLAS:2017svb} and an outer strip detector~\cite{ATLAS:2017azf}. It is designed to trigger in phase with the collision frequency of the HL-LHC, at \unit[40]{MHz} or every \unit[25]{ns}.

Modules form the base units of the ITk strip tracker~\cite{ATLAS:2020ize, Kuehn:2017cqo}. Sensors for each module have a strip pitch of \unit[$\sim$70]{\um} and an active depth of \unit[300]{\um}~\cite{Unno:2023bot}; glued onto each sensor are flex-circuits known as hybrids --- which instrument readout and control --- and powerboards~\cite{Haber:2023kha} --- which instrument high-voltage and monitoring. Mounted onto each hybrid are ATLAS Binary Chip (ABC) ASICs, and each ABC performs the readout of 256 strip channels. A measured charge above a set discriminator threshold for a given channel of the ABC results in a ``hit''. The production version of the ABC is known as the ABCStar~\cite{Cormier:2021oog}.

To ensure the ITk detector will meet its operational requirements in the radiation environment of the HL-LHC, the performance of strip modules and their electronics has been evaluated in numerous beam tests using electrons~\cite{Ruhr:2020qen, Arling:2023pio} as well as protons and heavy ions~\cite{Basso:2022qob, Dandoy:2023gpj, Belanger-Champagne:2023dqg}. From measurements performed with monochromatic photon beams~\cite{Poley:2016nwv, Blue:2021}, it has been observed that high fluxes of photons incident on a single strip --- $\unit[\mathcal{O}(1)]{photons/\unit[25]{ns}}$ --- cause the corresponding hit efficiency $\varepsilon$ to significantly deviate from the expected ``S-curve'':

\begin{equation}
    \varepsilon(q) = \frac{1}{2} \erfc\qty(\frac{q - q_\textrm{thr}}{\sqrt{2}\sigma}) \,,
    \label{eqn:scurve}
\end{equation}

\noindent where $q$ is the injected charge, $\erfc\qty(\ldots)$ is the complementary error function, $q_\textrm{thr}$ is the threshold charge, and $\sigma$ is the noise charge. In particular, too few and too many hits are measured at low and high values of $q_\textrm{thr}$, respectively, resulting in a broadening of the turn-on behaviour for that channel.

This effect is understood to be due to high occupancy of hits. The impulse response of the ABCStar's front-end (FE) extends well-beyond the \unit[25]{ns} readout, or ``trigger'', window of the detector. Therefore, hits in a given trigger window may affect hits in all subsequent trigger windows by effectively modulating the discriminator threshold, an effect coined ``threshold bounce''. At high occupancy, the ABCStar FE's output response is the superposition of its impulse response for each of the hits. In the case of the measurements performed with photon beams, the occupancy can reach 100\%, whereas ABCStar ASICs have been designed for a maximum occupancy of 10\%~\cite{Blue:2021}.

In this paper, a measurement of threshold bounce is reported using an ITk strip module in a monochromatic photon beam. The magnitude of the threshold bounce is studied for varying photon flux and discriminator thresholds, and a Monte Carlo (MC) program is developed to simulate the measured results. As part of this program, a mathematical formulation of the ABCStar FE's impulse response must be provided.

\subsection{The ABCStar FE’s impulse response}
\label{sec:impulse}

\begin{figure}[htbp]
    \centering
    \includegraphics[width=\textwidth]{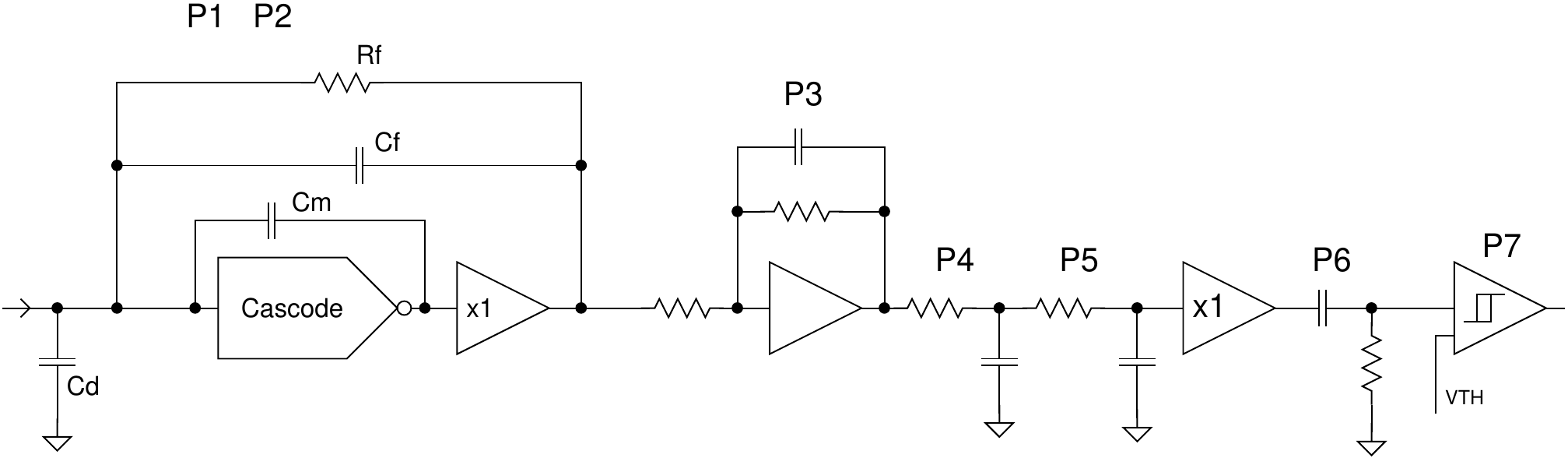}
    \caption{A simplified circuit diagram of the ABCStar's FE. The distinct blocks are described in the text of section~\ref{sec:impulse}, with time constant $\tau_i$ matched to the corresponding block denoted by ``P$i$'' for $i=1,\ldots,7$. As discussed in the text, the preamplifier (P1/P2) is resolved as two first-order low-pass filters within the semi-empirical approach. The first-order low-pass filter with time constant $\tau_7$ is found at the input to the discriminator (P7).}
    \label{fig:abcfe}
\end{figure}

A simplified circuit diagram of the ABCStar's FE is shown in figure~\ref{fig:abcfe}. It consists of a preamplifier, inverting amplifier, and shaper circuit, in that order; the shaper circuit consists of a second-order low-pass filter, a first-order high-pass filter, and a first-order low-pass filter, in that order. A semi-empirical model of the ABCStar's FE is used, where the preamplifier is resolved as two first-order low-pass filters in sequence.\footnote{This is a simplified approach, with the time constants for the two first-order low-pass filters chosen via tuning. A more correct treatment would follow the case of a cascode preamplifier with Miller compensation~\cite{Kaplon:2023yue}. However, for the purposes of this study, any functional form which closely follows simulation is suitable.} The impulse response\footnote{The impulse response of a system is defined as the time-domain output response of that system when injected with an input delta function.} of the semi-empirical model is given by:

\begin{equation}
    V_\textrm{ABCStar}(t) = V_0 \sum_{i=1}^7 \frac{C_i}{\tau_i}\exp\qty(\frac{-t}{\tau_i}) \,,
    \label{eqn:impulse_response}
\end{equation}

\noindent where $V_0$ corresponds to the absolute scale and:

\begin{itemize}
    \item $\tau_1$ ($= \unit[3]{ns}$) is the RC of preamplifier's first first-order low-pass filter;
    \item $\tau_2$ ($= \unit[14]{ns}$) is the RC of preamplifier's second first-order low-pass filter;
    \item $\tau_3$ ($= \unit[2]{ns}$) is the inverting amplifier's RC;
    \item $\tau_4$ ($= \unit[1.83]{ns}$) is the first pole of the second-order low-pass filter;
    \item $\tau_5$ ($= \unit[12.57]{ns}$) is the second pole of the second-order low-pass filter;
    \item $\tau_6$ ($= \unit[250]{ns}$) is the first-order high-pass filter's RC;
    \item $\tau_7$ ($= \unit[4.5]{ns}$) is the first-order low-pass filter's RC.
\end{itemize}

\noindent Here, $\qty{C_i \,;\, i=1,\ldots,7}$ are constants which depend on the time constants above. Details on the derivation of equation~\ref{eqn:impulse_response} are found in appendix~\ref{app:derivation}.

\begin{figure}[htbp]
    \centering
    \includegraphics[width=0.8\textwidth]{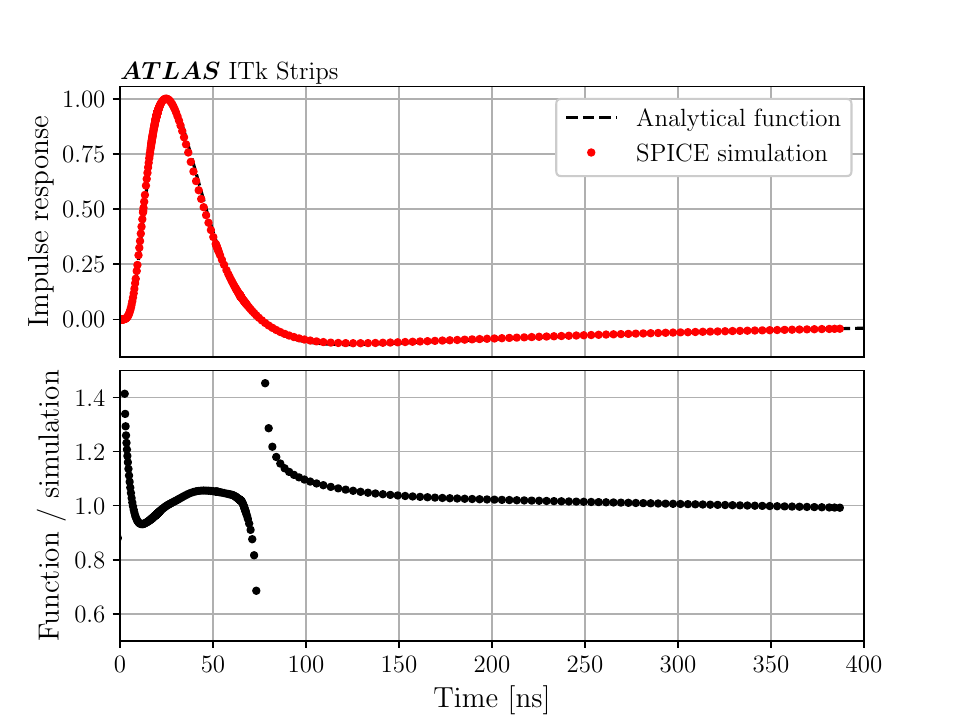}
    \caption{Impulse response of the ABCStar's FE. In the upper plot, both a SPICE simulation and an analytical function are shown are shown. The analytical function lies entirely underneath the SPICE simulation. In the lower plot, the ratio of the function to the simulation is shown. For the analytical function, the peaking time is \unit[25.6]{ns}, the full width at half maximum is \unit[33.2]{ns}, the zero crossing is \unit[74]{ns}, and the ratio of the positive-to-negative maxima is 8.8. The asymptotic behaviour in the ratio at 0 and \unit[74]{ns} is due to the numerical precision at these zero crossings.}
    \label{fig:impulse_response}
\end{figure}

Equation~\ref{eqn:impulse_response} is plotted alongside a SPICE simulation of the FE's impulse response in figure~\ref{fig:impulse_response}. The SPICE simulation was generated using \textsc{Spectre}~\cite{Spectre}, where the modelling of the circuit elements nearly corresponds to the filters described above. Good agreement is observed between the simulation and the analytical function. There are two regions of interest:

\begin{enumerate}
    \item $0 < t \lesssim \unit[75]{ns}$, where the impulse response is positive and expected to ``decrease'' the discriminator's threshold for subsequent trigger windows through constructive interference;
    \item $t \gtrsim \unit[75]{ns}$, where the impulse response is negative and expected to ``increase'' the discriminator's threshold for subsequent trigger windows through destructive interference. While the absolute maximum of this regime is nearly an order of magnitude smaller than that of the former, it is expected to affect many trigger windows given its long tail, $\unit[\mathcal{O}(100\textrm{s})]{ns}$.
\end{enumerate}

\section{Measurement setup}
\label{sec:setup}

The measurement was performed at the B16 Test Beamline~\cite{DLS} at the Diamond Light Source synchrotron on the Harwell Science and Innovation Campus in Didcot, United Kingdom. A Si (111) double-crystal monochromator~\cite{B16} provides a \unit[15]{keV} photon beam which is focused using an array of compound refractive lenses to a size of $\unit[2]{\um} \times \unit[3]{\um}$ at the detector plane.

\begin{figure}[htbp]
    \centering
    \begin{subfigure}{.49\textwidth}
        \centering
        \includegraphics[width=\textwidth]{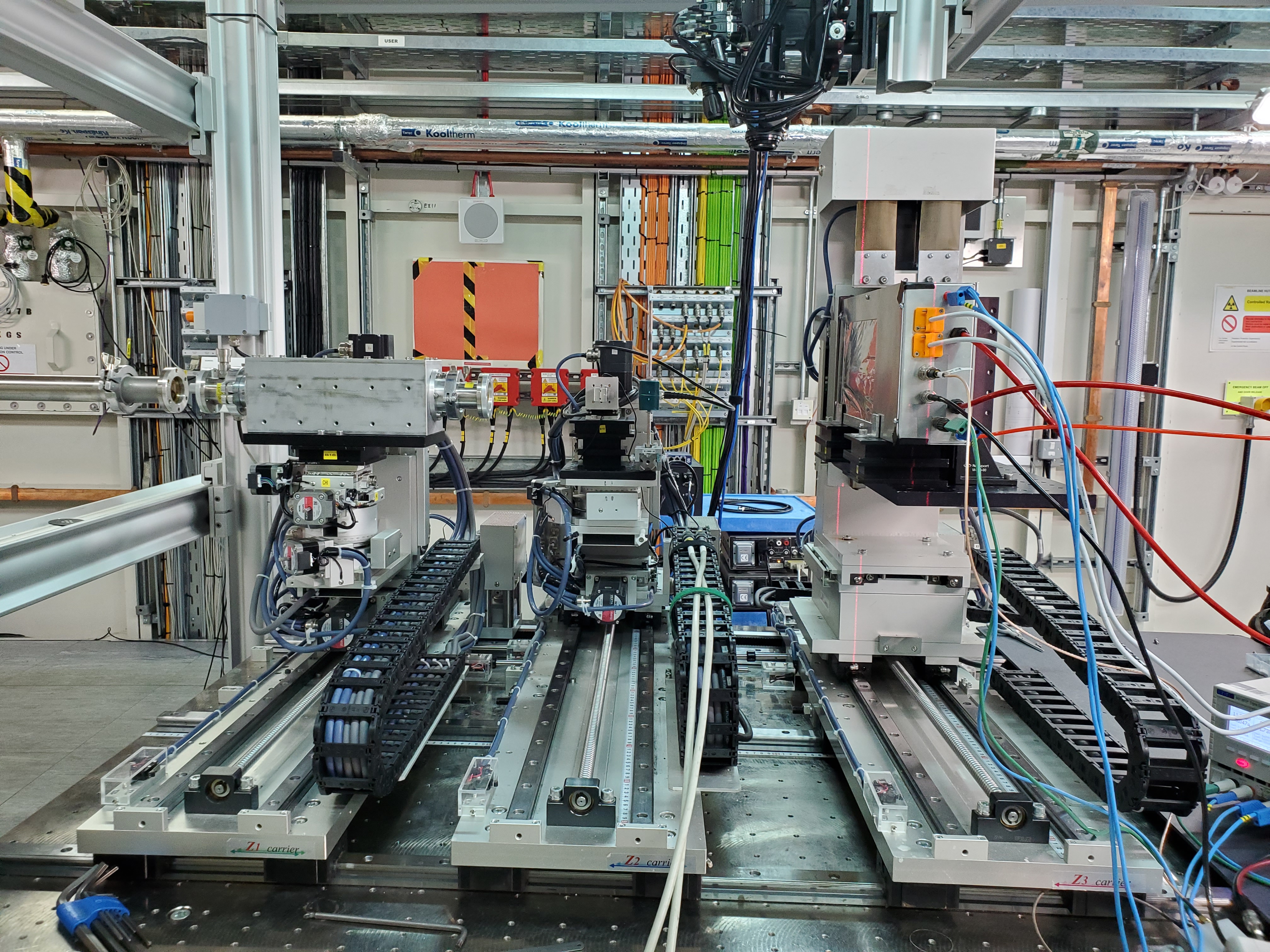}
        \caption{}
    \end{subfigure}
    \hfill
    \begin{subfigure}{.49\textwidth}
        \centering
        \includegraphics[width=\textwidth,angle=270]{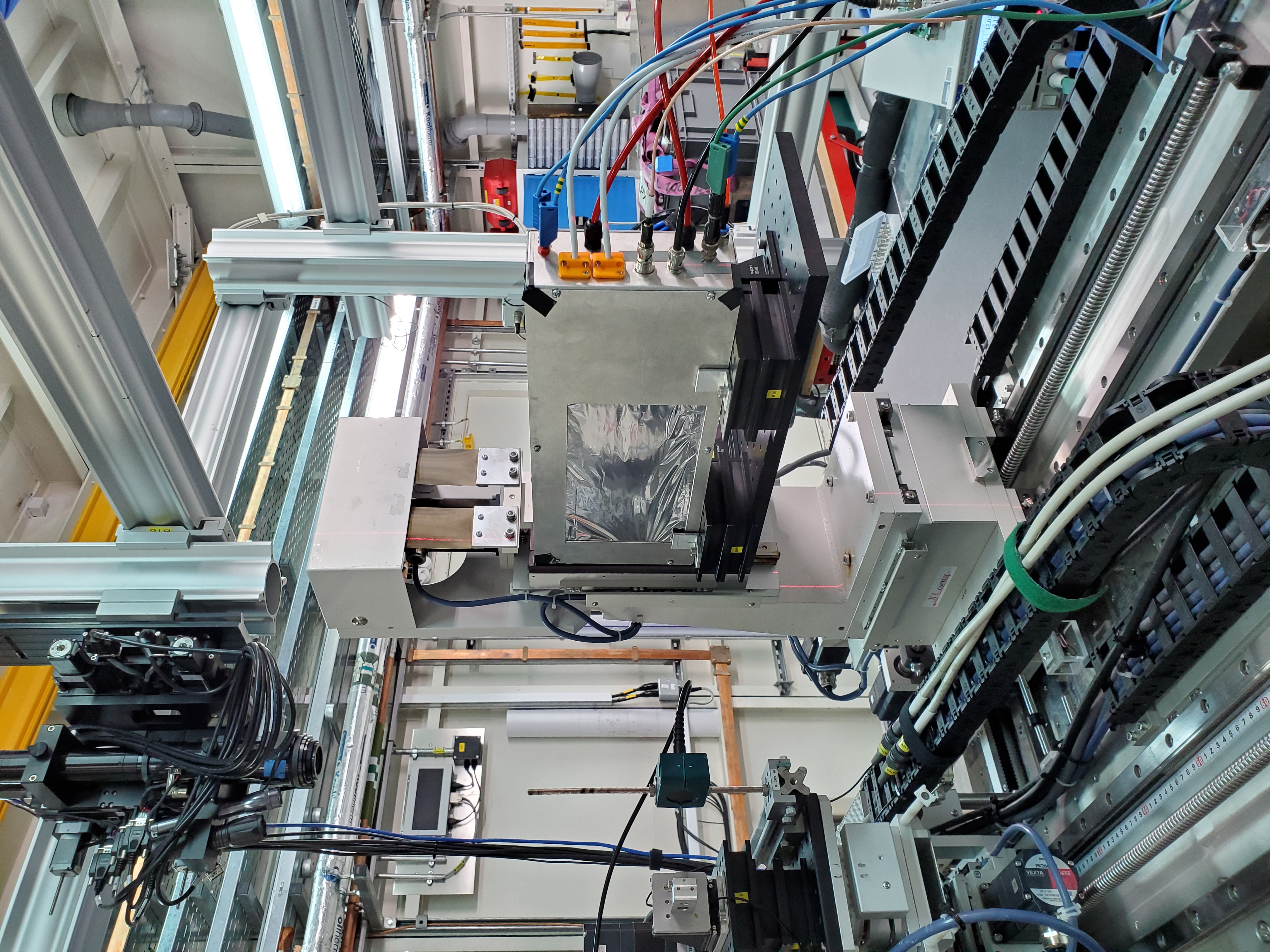}
        \caption{}
    \end{subfigure}
    \caption{A picture of the  measurement setup at the B16 Test Beamline at the Diamond Light Source. (a) A side view showing the beampipe (left), compound refractive lenses (centre left), and the module test box (centre right) mounted on precision translation stages (bottom). (b) An approximate $45^\circ$ view of the same setup.}
    \label{fig:setup}
\end{figure}

An ITk (``R5'') endcap strip module was mounted perpendicular to the beam inside a light-tight metal box. The silicon sensor was fully depleted by operating the module at a bias voltage of \unit[$-350$]{V}, and the box was flushed with nitrogen to guard against early sensor breakdown induced by high humidity~\cite{Fernandez-Tejero:2023asg}. A picture of the setup is shown in figure~\ref{fig:setup}. The measurement was performed at room temperature, corresponding to a temperature of \unit[$\sim$30]{\oC} on the module.

Precision translation stages were used to align the focused beam onto a single strip, corresponding to an incident photon flux of \unit[$\sim$0.10]{photons/ns} and measured using a calibrated photodiode. The choice of strip was fixed for all measurements. The active depth of the silicon sensor is \unit[300]{\um} at full depletion, and a \unit[15]{keV} photon has a 52\% probability of interacting within its volume. An interacting photon liberates a \unit[15]{keV} electron which, given the mean ionization energy of silicon is \unit[3.6]{eV}~\cite{Scholze:1996, Scholze:1998}, produces $\sim$4,200 electron-hole pairs while travelling up to \unit[20]{\um} within the silicon's volume. The corresponding input charge is \unit[0.67]{fC}, approximately $1/5$ of the charge deposited by a minimum ionizing particle and what the ABCStar FE was optimized for.\footnote{The difference between the input charge collected during operation and that for which the ABCStar FE was optimized is less drastic at the ITk's end-of-life, where the charge collection efficiency of the silicon is reduced.} This charge is deposited entirely within the volume of a single strip; there is no charge sharing with neighbouring strips.

The module was interfaced with a Nexys Video FPGA board~\cite{DigilentNV}, and read out using the Inner Tracker Strips Data Acquisition (ITSDAQ) software~\cite{ATLAS:2020ize} and firmware~\cite{ITSDAQfw}.

To explore the impact of threshold bounce for varying particle flux, the photon flux was decreased from the machine flux by placing aluminum attenuators of known thickness between the source of the beam and the module. Photon flux and beam intensity are used synonymously for the remainder of the paper, the latter measured relative to the machine flux (i.e., \unit[0.1]{photons/ns}). For each flux, the discriminator threshold was varied between 0 and 60 digital-to-analog converter (DAC) counts; this covers the full range of input charge expected during operation, where \unit[15]{DAC} corresponds to an expected hit efficiency of 50\% for the nominal photon signal (see figure~\ref{fig:norm_hits}). For each threshold, groups of 128 triggers were sent --- \unit[25]{ns} apart (i.e., the LHC bunch-crossing period) and asynchronous to the beam structure --- and the entire module was read out. Groups were iteratively sent until some total number of triggers was reached. As the photon flux became smaller, the total number of triggers was increased in order to measure the number of hits with sufficient precision.

\section{Results}
\label{sec:results}

\begin{figure}[htbp]
    \centering
    \includegraphics[width=0.7\textwidth]{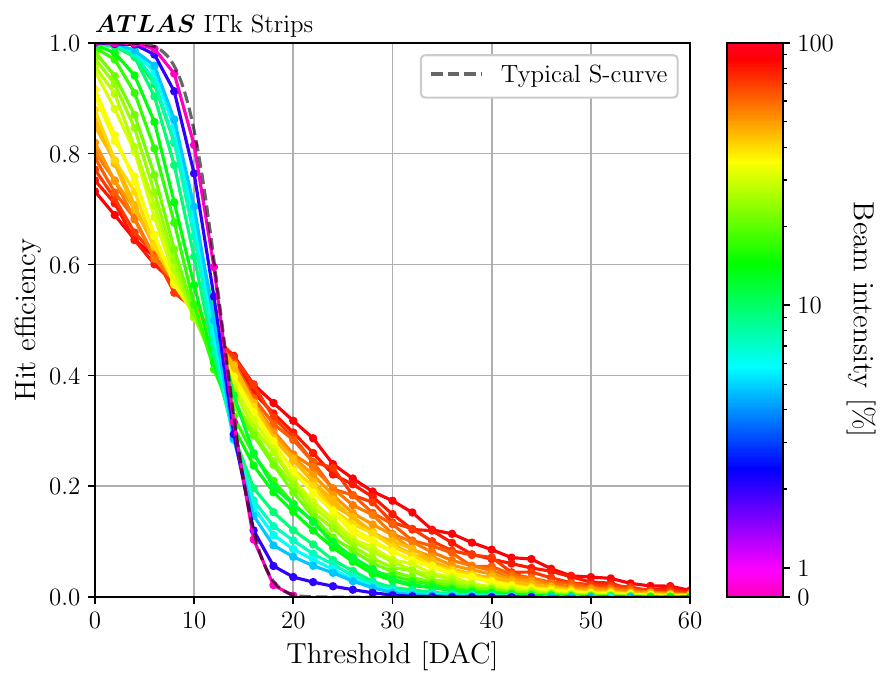}
    \caption{The hit efficiency as a function of the discriminator threshold and the beam intensity. The shape of a typical S-curve is overlaid for comparison. The colourbar uses a logarithmic scale except in the range of $[0, 1]\%$, where a linear scale is used instead.}
    \label{fig:norm_hits}
\end{figure}

For each photon flux and discriminator threshold, the total number of measured hits were counted and normalized to the total number of triggers sent for that particular flux and threshold. Figure~\ref{fig:norm_hits} shows the hit efficiency as a function of each parameter. In agreement with the observations of reference~\cite{Blue:2021}, the hit efficiency is reduced for high photon fluxes, increasing in efficiency and approaching the standard S-curve shape as the flux is reduced.

The effect of threshold bounce is expected to be most pronounced in the fraction of triggers which measure a hit as a function of the number of trigger windows or ``distance'' following a hit. The distance is counted up to 127 trigger windows away, as triggers are sent in groups of 128. As an example, consider a group of 18 triggers with the corresponding hit decisions:

\begin{equation}
    0 \, 1 \, 1 \, 0 \, 0 \, 1 \, 0 \, 1 \, 1 \, 0 \, 0 \, 1 \, 0 \, 0 \, 0 \, 1 \, 1 \, 0 \,.
    \label{eqn:hit_sequence}
\end{equation}

\noindent The calculation of the fraction of triggers resulting in a hit starts at the first hit of the group of 18 triggers, resetting each time a hit is encountered. By ``unravelling'' the sequence of hits above, the calculation proceeds as shown in table~\ref{tab:example}. In the table, there is always a hit at a distance of 0 following a hit (i.e., itself). The numerator is calculated as the number of hits found a distance $i$ following a hit (i.e., the number of 1s in a column) and the denominator is calculated as how often a distance $i$ is reached following a hit (i.e., the number of 1s and 0s in a column). The fraction is simply the numerator divided by the denominator. Each trigger following the last hit of a group is included in the calculation as well. If both the numerator and the denominator are zero, the fraction is set to zero.

\begin{table}[htbp]
    \centering
    \caption{Example calculation of the fraction of triggers which measure a hit as a function of the distance following a hit. The example applies to the hit decisions in equation~\ref{eqn:hit_sequence}.}
    \label{tab:example}
    \begin{tabular}{| r | c c c c c c c c |}
        \toprule
        & \multicolumn{8}{|c|}{Distance from hit [\# of \unit[25]{ns} steps]} \\
        & 0 & 1 & 2 & 3 & 4 & 5 & \ldots & 17 \\
        \midrule
        \multirow{8}{*}{Hit decision} & 1 & 1 & -- & -- & -- & -- & \ldots & -- \\
        & 1 & 0 & 0 & 1 & -- & -- & \ldots & -- \\
        & 1 & 0 & 1 & -- & -- & -- & \ldots & -- \\
        & 1 & 1 & -- & -- & -- & -- & \ldots & -- \\
        & 1 & 0 & 0 & 1 & -- & -- & \ldots & -- \\
        & 1 & 0 & 0 & 0 & 1 & -- & \ldots & -- \\
        & 1 & 1 & -- & -- & -- & -- & \ldots & -- \\
        & 1 & 0 & -- & -- & -- & -- & \ldots & -- \\
        \midrule
        Numerator & 8 & 3 & 1 & 2 & 1 & 0 & \ldots & 0 \\
        Denominator & 8 & 8 & 4 & 3 & 1 & 0 & \ldots & 0 \\
        \midrule
        Fraction & 1.000 & 0.375 & 0.250 & 0.667 & 1.000 & 0.000 & \ldots & 0.000 \\
        \bottomrule
    \end{tabular}
\end{table}

The calculation above is naturally extended to groups of 128 triggers. The uncertainty on the fraction is calculated using the 68\% Wilson confidence interval~\cite{Wilson:1927}, which has the benefit of being nonzero even when the numerator is zero.

\begin{figure}[htbp]
    \centering
    \includegraphics[width=0.7\textwidth]{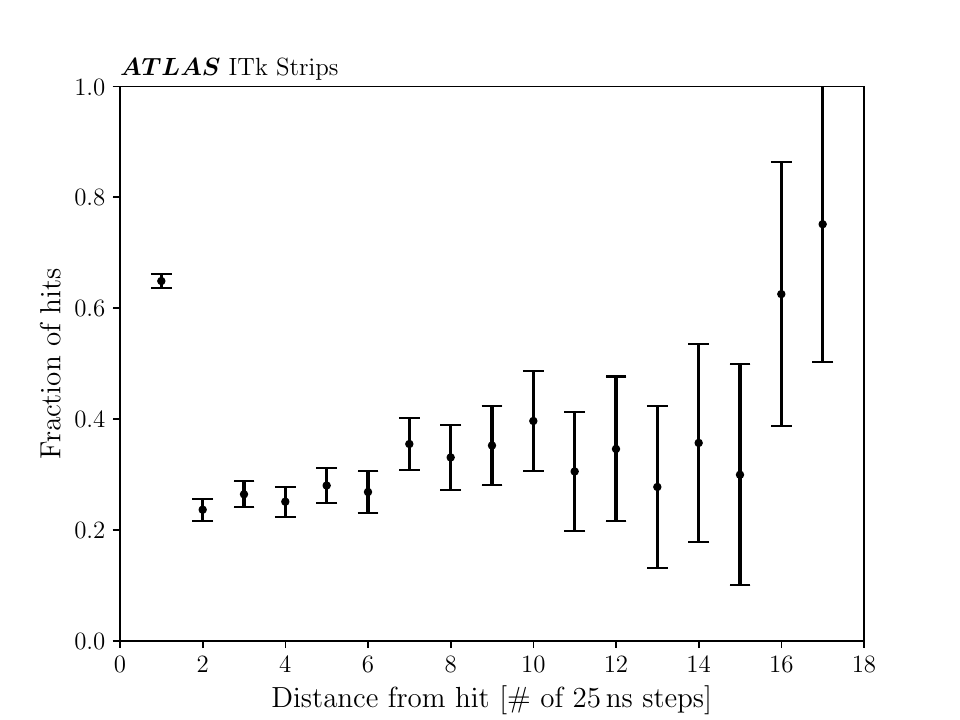}
    \caption{The observed fraction of triggers which measure a hit as a function of the distance following a hit for a beam intensity of 89\% and a discriminator threshold of \unit[14]{DAC}.}
    \label{fig:frac_obs_run225_thr12}
\end{figure}

Because of the shape of the impulse response, it is expected that the trigger immediately following a hit (i.e., \unit[25]{ns} later) will have a higher probability of measuring a hit as the positive-valued impulse response lowers the effective discriminator threshold; in contrast, triggers subsequent to the trigger immediately following a hit (i.e., \unit[50]{ns}, \unit[75]{ns}, $\ldots$ later) will have lower probability of measuring a hit as the negative-valued impulse response increases the effective threshold. As the impulse response asymptotically approaches zero, the probability of measuring a hit will increase. This behaviour is demonstrated in figure~\ref{fig:frac_obs_run225_thr12}, where the observed fraction of hits versus distance from a hit is shown for an example beam intensity and discriminator threshold.

In order to estimate the ``magnitude'' of the threshold bounce effect for a given flux and threshold, a linear fit is performed to the fraction versus distance for distances greater than 2. Only data points with at least one denominator event are considered. The magnitude of the threshold bounce effect is assumed to be proportional to the magnitude of the slope. If there was no threshold bounce, the slope would be consistent with zero; in other words, all triggers following a hit have an equal probability (i.e., fraction) of measuring a hit as they are independent of one another.

\begin{figure}[htbp]
    \centering
    \includegraphics[width=\textwidth]{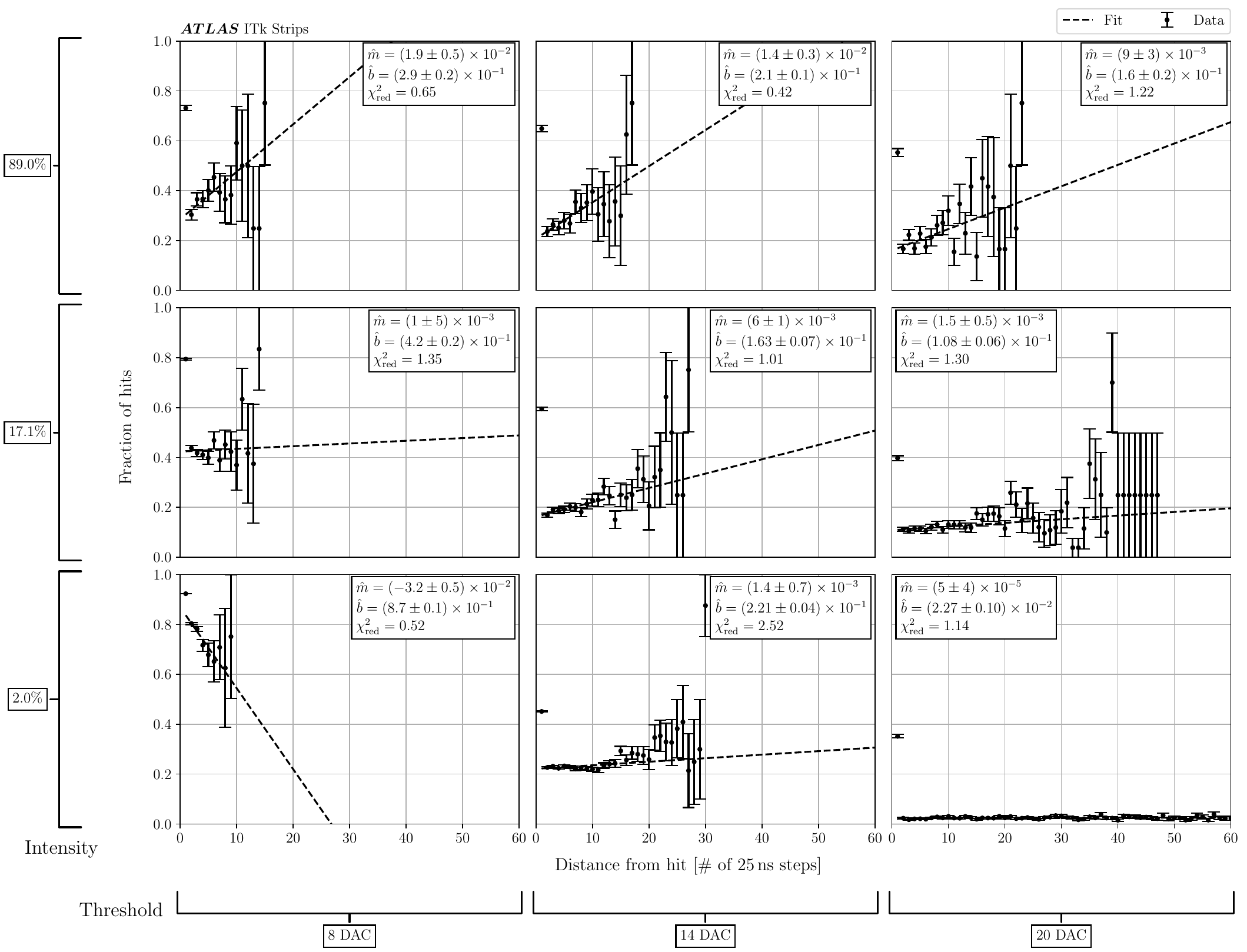}
    \caption{The observed fraction of triggers which measure a hit as a function of the distance following a hit for nine combinations of beam intensities (89\%, 17.1\%, and 2\%) and discriminator thresholds (8, 14, and \unit[20]{DAC}). Also shown are the linear fits to the observed data. The error bars correspond to the Wilson confidence intervals.}
    \label{fig:fits_3x3}
\end{figure}

\begin{figure}[htbp]
    \centering
    \begin{subfigure}{0.7\textwidth}
        \centering
        \includegraphics[width=\textwidth]{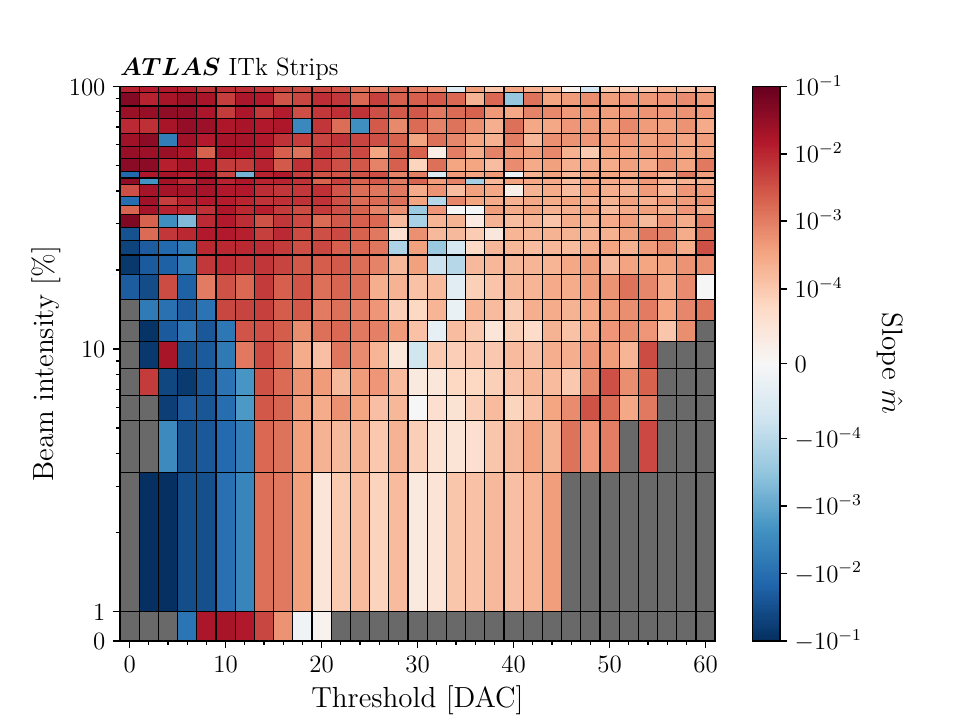}
        \caption{Slope}
    \end{subfigure} \\
    \begin{subfigure}{0.7\textwidth}
        \centering
        \includegraphics[width=\textwidth]{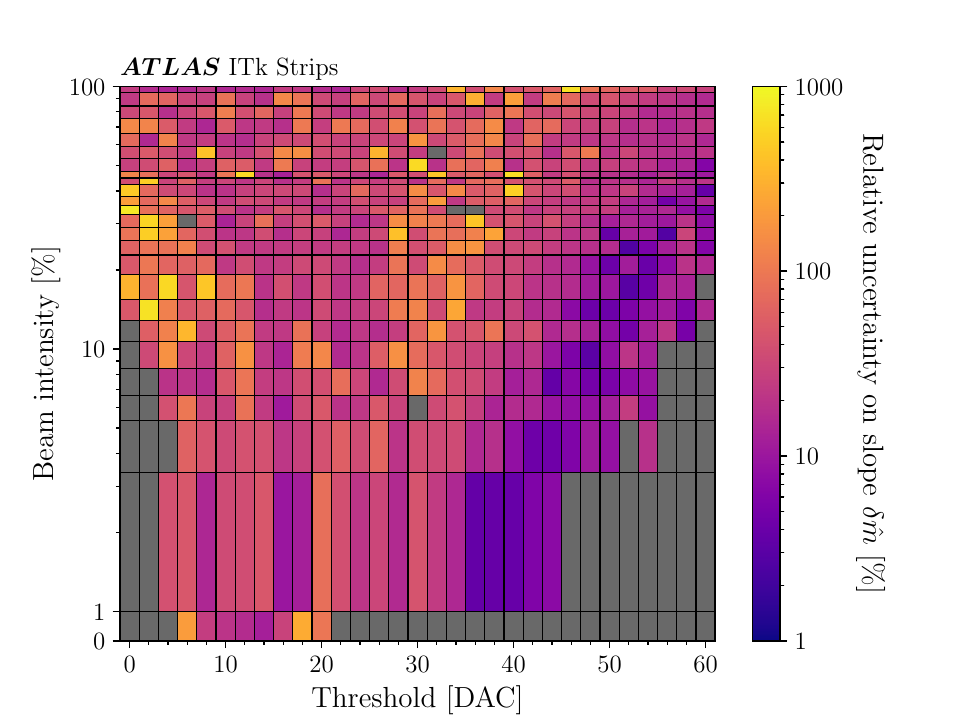}
        \caption{Relative uncertainty on slope}
    \end{subfigure}
    \caption{The results of the linear fits to the fraction of triggers which measure a hit as a function of the distance following a hit for all fluxes and thresholds, including (a) the fitted slopes and (b) the corresponding uncertainties on the slopes. The grey cells indicate the fluxes/thresholds where there is not enough data to perform a fit or where the associated slope or uncertainty values go off the scale of the colourbar, also due to a lack of data. In (a), the colourbar uses a logarithmic scale except in the range of $[-10^{-4}, 10^{-4}]$, where a linear scale is used instead. In both (a) and (b), the $y$-axis uses a logarithmic scale except in the range of $[0, 1]\%$, where a linear scale is used instead.}
    \label{fig:slope_and_unc}
\end{figure}

\begin{figure}[htbp]
    \centering
    \begin{subfigure}{0.7\textwidth}
        \centering
        \includegraphics[width=\textwidth]{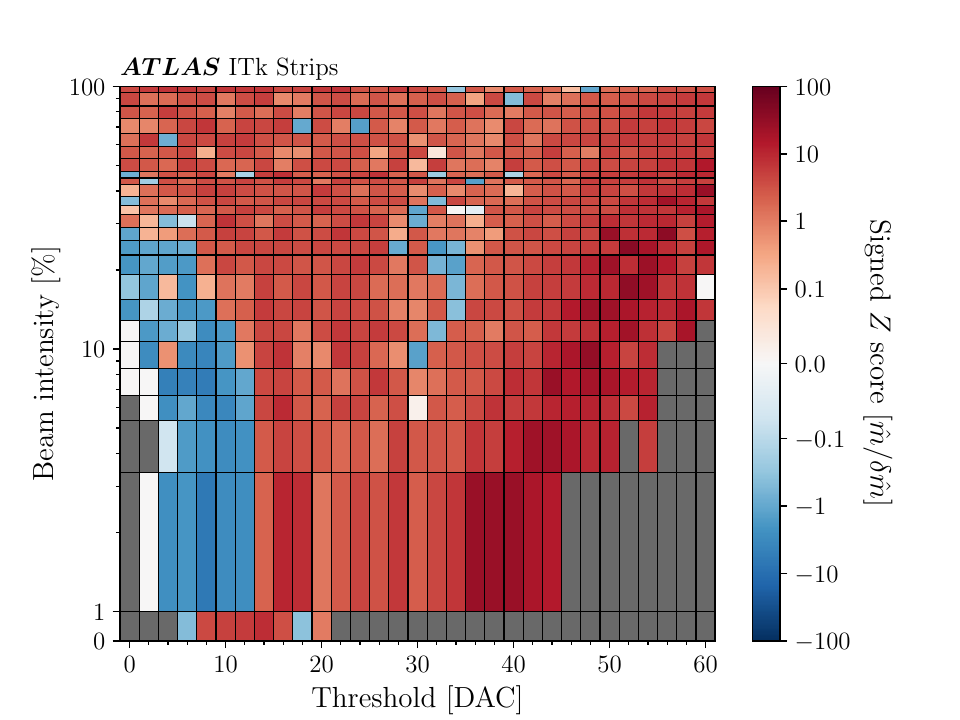}
        \caption{Signed $Z$ score}
    \end{subfigure} \\
    \begin{subfigure}{0.7\textwidth}
        \centering
        \includegraphics[width=\textwidth]{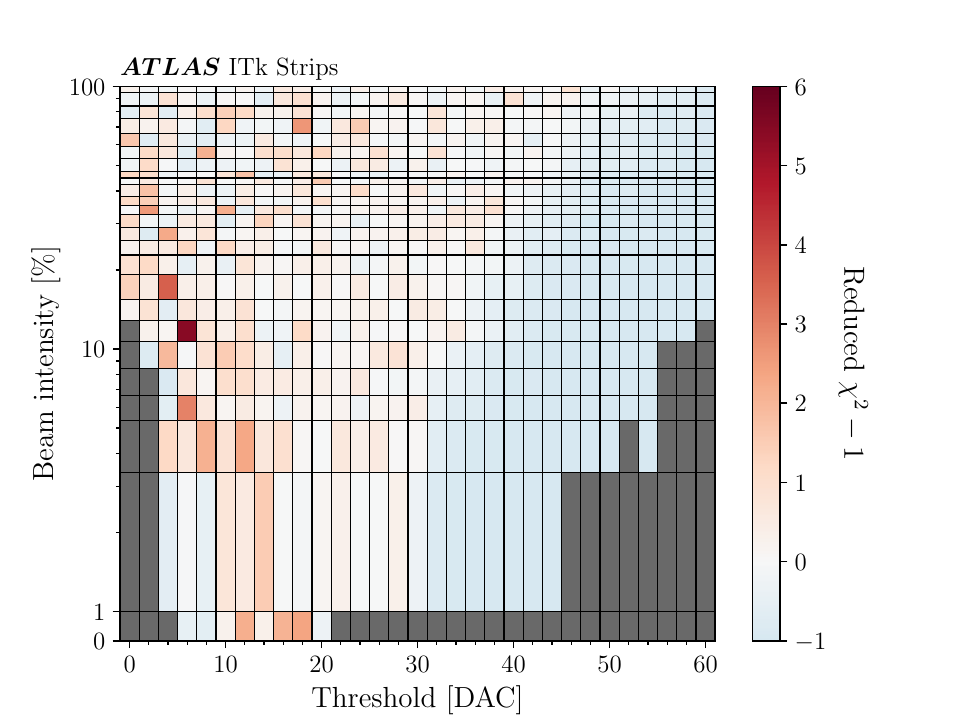}
        \caption{Reduced $\chi^2 - 1$}
    \end{subfigure}
    \caption{The results of the linear fits to the fraction of triggers which measure a hit as a function of the distance following a hit for all fluxes and thresholds, including (a) the signed $Z$ scores and (b) the corresponding values of the reduced $\chi^2 - 1$. The grey cells indicate the fluxes/thresholds where there is not enough data to perform a fit or where the associated $Z$ scores or $\chi^2$ values go off the scale of the colourbar, also due to a lack of data. In (a), the colourbar uses a logarithmic scale except in the range of $[-0.1, 0.1]$, where a linear scale is used instead. In both (a) and (b), the $y$-axis uses a logarithmic scale except in the range of $[0, 1]\%$, where a linear scale is used instead.}
    \label{fig:zscore_and_chisq}
\end{figure}

The observed data and the resulting linear fits for a number of example beam intensities and discriminator thresholds are shown in figure~\ref{fig:fits_3x3}. Figures~\ref{fig:slope_and_unc} and \ref{fig:zscore_and_chisq} show the important results of the linear fits for all intensities and thresholds: slope, uncertainty on the slope, $Z$ score (``significance''), and reduced $\chi^2$ ($\chi^2_\textrm{red}$). There are three regions of note:

\begin{enumerate}
    \item $\textrm{Intensity} \in [30, 100]\%$ and $\textrm{threshold} \in [10, 30]\,\unit{DAC}$: there is a clear region at high fluxes and moderate thresholds where the slope is nonzero and positive ($\mathcal{O}(10^{-2}$)) with good precision ($\mathcal{O}(10\%)$), leading to $Z$ scores $\mathcal{O}(1)$. Additionally, the fits adequately describe the data in this region ($\chi^2_\textrm{red} \sim 1$). This is the region where the effect of threshold bounce is strongest and results in less and more measured hits at low and high thresholds, respectively;
    \item $\textrm{Threshold} \gtrsim \unit[30]{DAC}$ (and decreasing for decreasing flux): at high thresholds, the slope approaches zero from the positive direction ($\lesssim$$10^{-4}$) with high precision ($\lesssim$10\%), leading to high $Z$ scores ($\gtrsim$10). However, the linear models over-fit the data in this region ($\chi^2_\textrm{red} \sim 0$), indicating that the data uncertainties may be overestimated. The results are sensible: as the flux decreases, the distance between subsequent photons increases, reducing the probability of one hit influencing another;
    \item $\textrm{Intensity} \lesssim 30\%$ and $\textrm{threshold} \lesssim \unit[10]{DAC}$: at low fluxes and low thresholds, the slope is nonzero and negative ($\mathcal{O}(-10^{-2})$). This effect is believed\footnote{Compared to regions~1 and 2, the behaviour in region~3 is more complicated and therefore difficult to understand completely.} to be a result of gaps in the beam structure: at these low thresholds, a photon will almost always result in a hit; however, with a photon flux of \unit[$<$1]{photon/\unit[25]{ns}}, subsequent photons will be spaced \unit[$>$25]{ns} apart and the effective threshold will always be enhanced by the negative tail of the ABCStar FE's impulse response. As the set threshold increases, groups of photons which have pushed the effective threshold higher and higher become ``filtered'' away, resulting in behaviour similar to that at higher fluxes. It is worth noting that the linear fits are somewhat poorer in this region ($\chi^2_\textrm{red} > 1)$ as compared to region~1.
\end{enumerate}

\noindent Overall, the measured results agree with physical intuition.

At the HL-LHC, the expected channel occupancy is $\sim$1\%, and incident charged particles are expected to liberate \unit[$\sim$1]{fC} of charge as they pass through a module's silicon sensor~\cite{ATLAS:2017azf}. The 2\% intensity and \unit[20]{DAC} threshold subplot of figure~\ref{fig:fits_3x3} is most representative of these conditions. For an incident photon flux of \unit[$\mathcal{O}(1)$]{photons/\unit[25]{ns}}, 98\% attenuation leads to a channel occupancy of 2\%, as every photon results in a hit for low enough discriminator thresholds. Additionally, a DAC threshold of 20 corresponds to an approximate \unit[1]{fC} charge threshold.\footnote{This is apparent from the response curve of our tested module, shown later in figure~\ref{fig:response_curve}.} For these conditions, the detector will operate in the ``flat'' region, where the fraction of hits versus distance from a hit is nearly constant (beyond the initial enhancement at distance = 1). This is apparent from the aforementioned subplot.

In figures~\ref{fig:slope_and_unc} and \ref{fig:zscore_and_chisq}, it is worth highlighting that there are grey cells for which the results of the fits are not presented. These grey cells correspond to scenarios where there is not enough data to perform the fit: in the associated fraction versus distance data, only one or two distance values have nonzero fractions, leading to undefined linear fits or linear fits with zero covariance in the fitted parameters, respectively. Therefore, results are not shown for these edges cases.

\FloatBarrier

\subsection{Simulation}
\label{sec:simulation}

In order to validate the physical mechanism of threshold bounce, a Monte Carlo simulation was developed to compare with the measured data. The simulation assumes the charge injected by a photon follows a delta function\footnote{This is a simplification: the injected charge will have a nonzero width in the time domain, and so the output response of the ABCStar FE will be the convolution of the charge's time-domain behaviour with the FE's impulse response. The convolution translates to a smeared output response, modifying its peaking time and extending its negative tail to later times. The latter effect may lead to enhanced threshold bounce.} and proceeds as follows:

\begin{enumerate}
    \item In time steps of $0 < \dt \equiv (t_{i+1} - t_i) \leq \unit[25]{ns}$, photons are drawn from a Poisson distribution with a mean $\mu = \phi \times \dt$, where $\phi$ is the photon flux. In this paper, $\phi = \unit[0.10]{photons/ns}$ is assumed and $\dt = \unit[1]{ns}$ is chosen;
    \item Photons are attenuated according to the thickness of the aluminum attenuator (with an attenuation coefficient of $(\unit[465]{\um})^{-1}$ for a photon energy of \unit[15]{keV}~\cite{NIST}) and required to interact within the \unit[300]{\um} active depth of the silicon sensor (with an attenuation coefficient of $(\unit[415]{\um})^{-1}$ for a photon energy of \unit[15]{keV}~\cite{NIST}). In either case, the probability of survival is given by $\exp(-\mu\times\ell)$, where $\mu$ is the attenuation coefficient and $\ell$ is the thickness of the material;
    \item Starting from the trailing edge of that time bin, $t_{i+1}$, the impulse response of the ABCStar, equation~\ref{eqn:impulse_response}, is evaluated for all future times and added to the existing response of the ABCStar (at $t = 0$, the response is simply the pedestal). The scale of the impulse response is proportional to the number of photons in the corresponding time bin;
    \item Steps 1 through 3 are repeated until a trigger edge, $t_{i+1}\,\textrm{mod}\,(\unit[25]{ns}) = 0$, is reached (excluding $t = 0$). At this time, the total response of the ABCStar is summed with zero-centred Gaussian output noise and compared to the set discriminator threshold. If the sum exceeds the threshold, a hit is recorded for that trigger window;
    \item Steps 1 through 4 are repeated until 128 triggers are reached. At this time, the detector undergoes a ``reset'' whereby the response of the ABCStar is returned to its $t = 0$ state;
    \item Steps 1 through 5 are repeated until the number of iterations in simulation matches that in data; in other words, until we have simulated as many events as were recorded for each beam intensity.
\end{enumerate}

\begin{figure}[htbp]
    \centering
    \includegraphics[width=\textwidth]{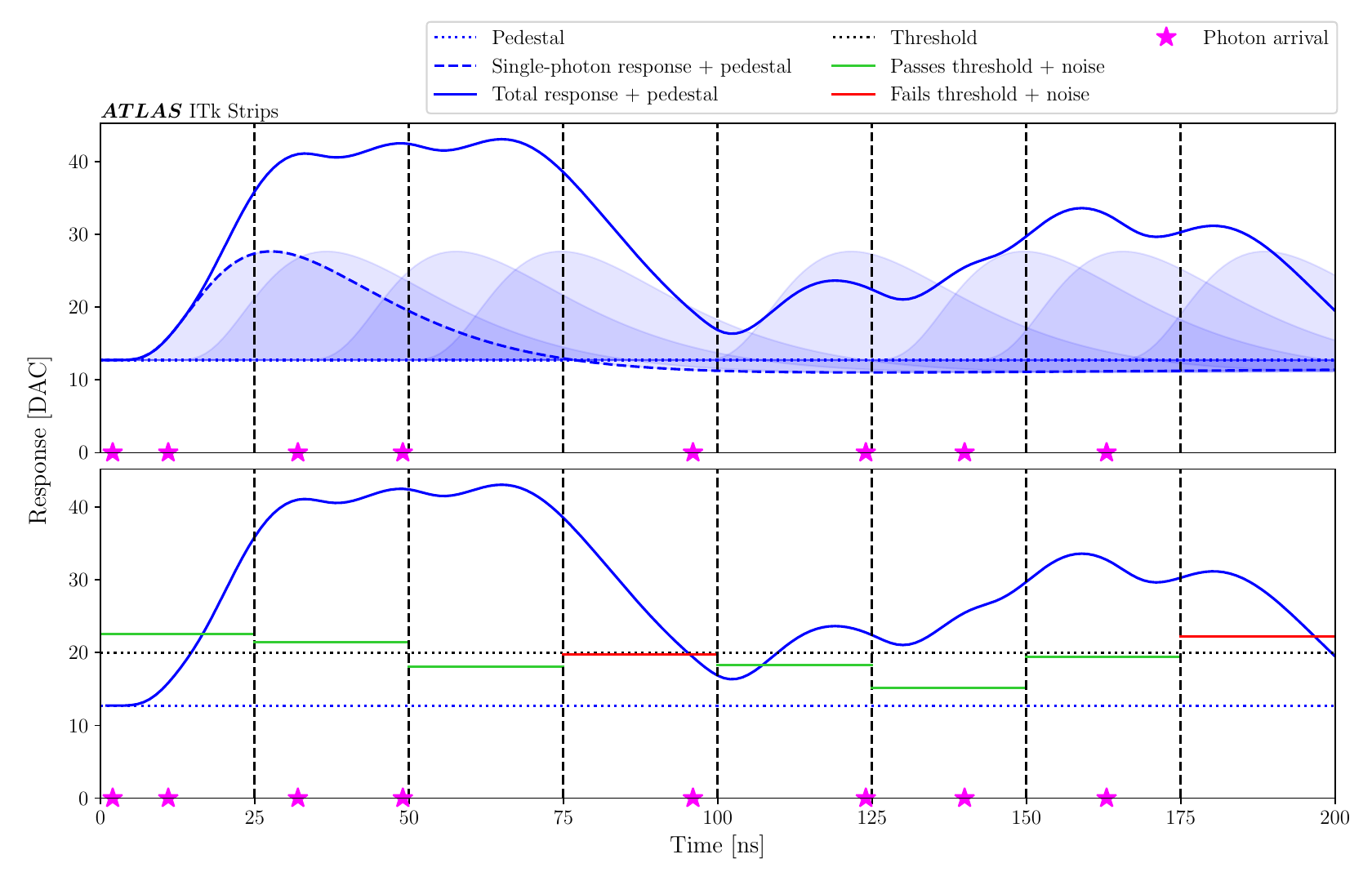}
    \caption{A demonstration of the ABCStar FE's output response for eight simulated events (i.e., triggers) at a beam intensity of 90\% and a discriminator threshold of \unit[20]{DAC}. The magenta stars indicate the arrival of photons. In the upper plot, the impulse response, equation~\ref{eqn:impulse_response}, is drawn at each of these times as a shaded blue area. The dashed blue line is drawn to emphasize the impulse response for the first photon. The solid blue line represents the total FE response, given by the linear sum of the impulse response for each photon across all time bins. All of these quantities are defined relative to the pedestal, which is indicated by the dotted blue horizontal line. N.B. the long tail of the impulse response which falls below the pedestal. In the lower plot, the total FE response is shown in relation to the relevant trigger thresholds. The dashed black vertical lines indicate trigger times (i.e., every \unit[25]{ns}). The dotted black horizontal line indicates the set threshold; the solid green horizontal lines indicate the set threshold plus the noise if the total FE response exceeds this sum at a trigger time (i.e, a hit), and the solid red horizontal lines indicate this same quantity if the total FE response does not exceed this sum at a trigger time (i.e., no hit).}
    \label{fig:simulation}
\end{figure}

\noindent The output response of the ABCStar across several simulated trigger windows is shown in figure~\ref{fig:simulation}, visually demonstrating steps 1 through 4 above. In particular, it highlights the \unit[25]{ns} delay between when charge is injected and when the FE response peaks --- a result of the FE's impulse response, equation~\ref{eqn:impulse_response} --- and demonstrates the total FE response to be the linear sum of the impulse response for each independent charge injection (for example, due to the arrival of a photon).

\begin{figure}[htbp]
    \centering
    \includegraphics[width=0.8\textwidth]{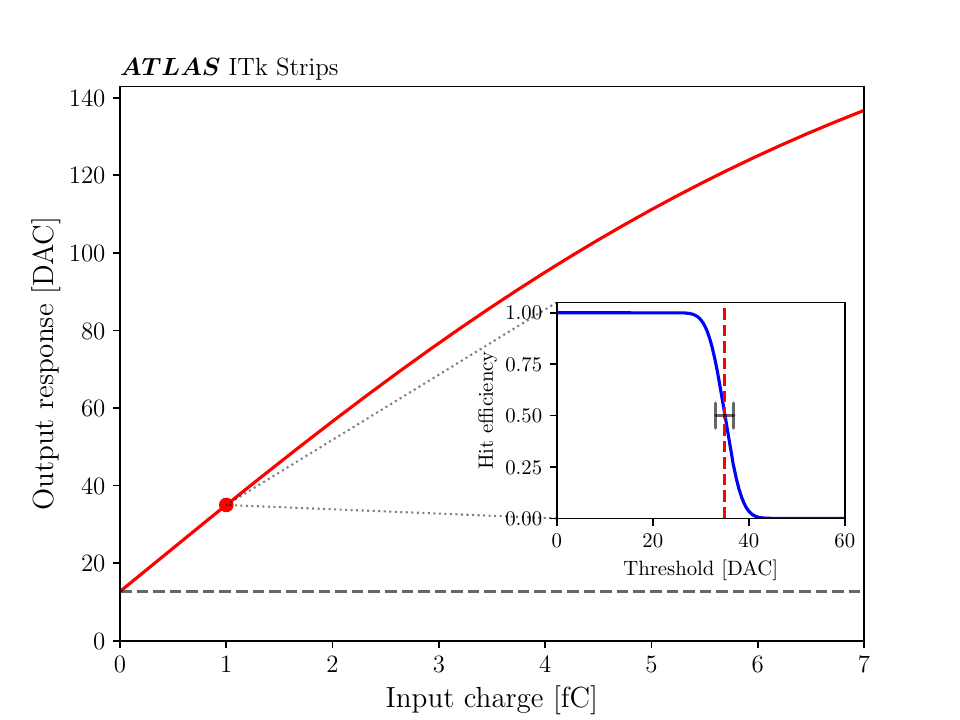}
    \caption{The response curve used in the simulation. The gain at an input charge $q$ corresponds to the slope of the response curve at $q$. The value of each point in the response curve corresponds to value of the set threshold which results in a 50\% hit efficiency, determined by scanning the hit efficiency as a function of the set threshold and fitting equation~\ref{eqn:scurve}, as shown in the inset. The fitted mean corresponds to the value of the response curve and the fitted width corresponds to the value of the output noise.}
    \label{fig:response_curve}
\end{figure}

The ABCStar's pedestal, the absolute scale of its impulse response, and the detector noise are measured using the full-test suite of the ITk strip modules~\cite{ATLAS:2020ize}, which includes measurements of each channel's trim, strobe delay, gain, and input noise --- some of these quantities are shown in figure~\ref{fig:response_curve}. As these characteristics are temperature-dependent and because the conditions of the full-test did not necessarily match the conditions of the threshold bounce measurements, scale factors $\mathcal{O}(1)$ are applied to each to improve the agreement between data and simulation. For the simulation above, the pedestal is taken to be \unit[12.7]{DAC} (\unit[35.7]{mV}), the absolute scale of the impulse response is taken to be \unit[22.3]{DAC/fC} (\unit[61.8]{mV/fC}), and the output noise is taken to be \unit[2.8]{DAC} (\unit[7.6]{mV}), where the bracketed quantities include the corresponding DAC-to-mV conversion.

To assess the impact of the \emph{absence} of threshold bounce on the simulation, the impulse response, equation~\ref{eqn:impulse_response}, was set to zero for times greater than \unit[25]{ns}:

\begin{equation}
    V^\prime_\textrm{ABCStar}(t) = \begin{cases}
        V_\textrm{ABCStar}(t) & t \leq \unit[25]{ns} \\
        0 & t > \unit[25]{ns} \\
    \end{cases} \,.
    \label{eqn:impulse_response_no_tb}
\end{equation}

\noindent Doing so ensures that the shape of the impulse response is correctly used to determine the hit decision for photons arriving within a particular trigger window but not subsequent trigger windows. The simulation was repeated using the modified impulse response $V^\prime_\textrm{ABCStar}(t)$.

\begin{figure}[htbp]
    \centering
    \includegraphics[width=\textwidth]{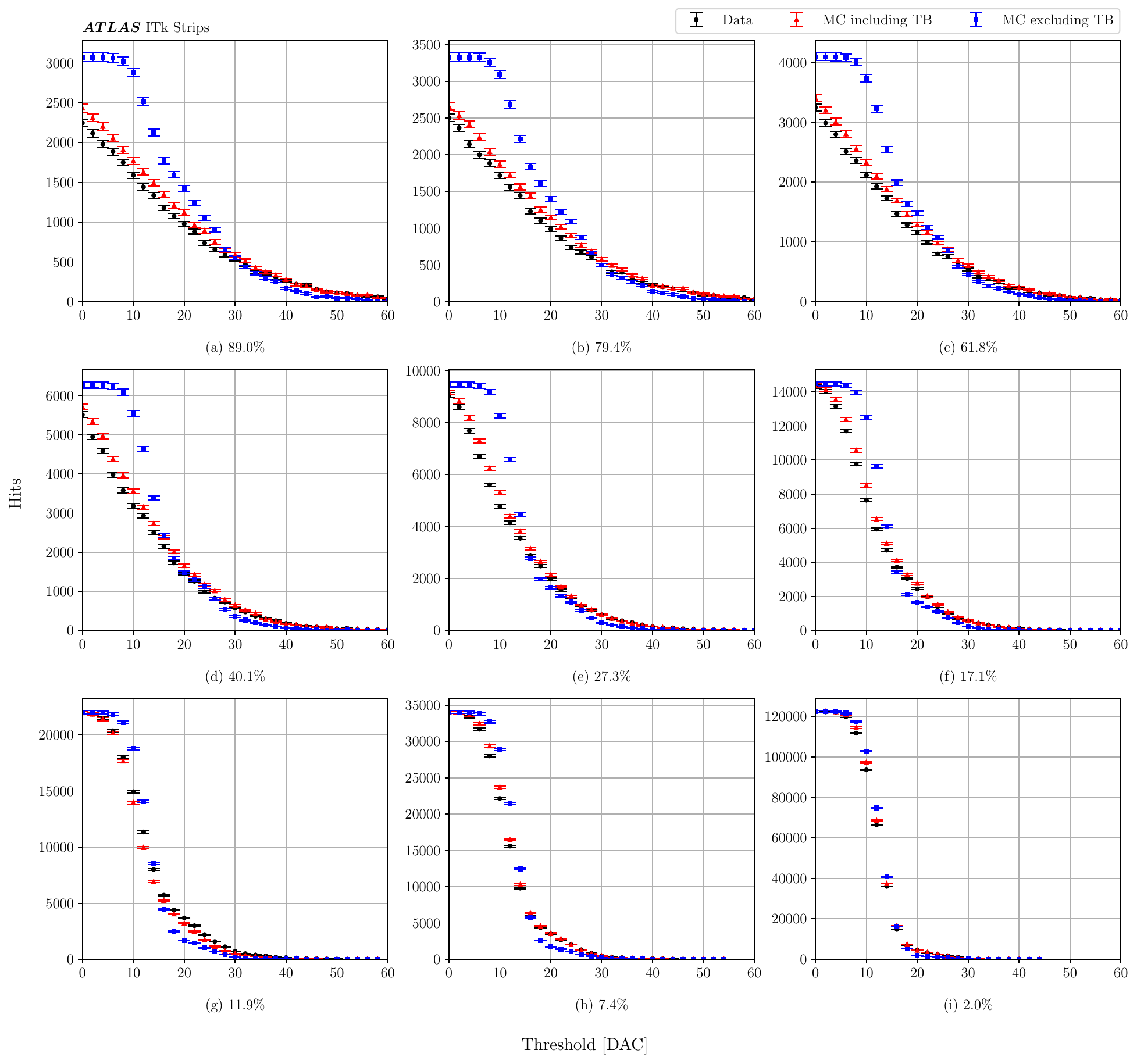}
    \caption{The number of observed and simulated hits as a function of threshold for nine different beam intensities. Simulations including and excluding threshold bounce (``TB'') effects are both shown. The error bars correspond to the Poisson counting uncertainties.}
    \label{fig:nhits_sim_vs_obs_3x3}
\end{figure}

\begin{figure}[htbp]
    \centering
    \begin{subfigure}{0.7\textwidth}
        \centering
        \includegraphics[width=\textwidth]{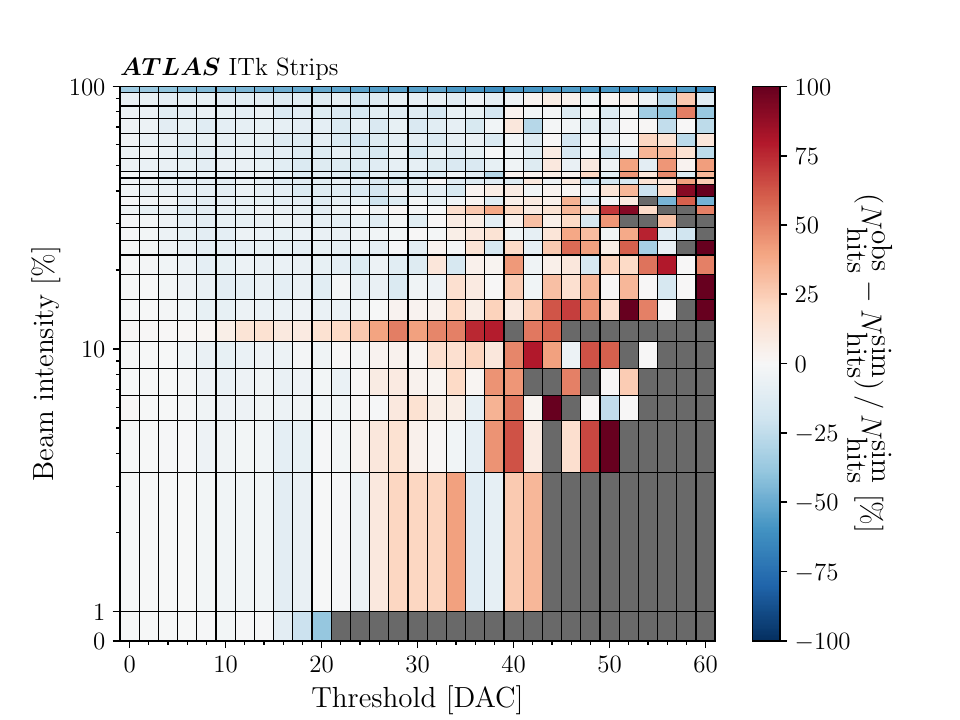}
        \caption{Including threshold bounce effects}
    \end{subfigure} \\
    \begin{subfigure}{0.7\textwidth}
        \centering
        \includegraphics[width=\textwidth]{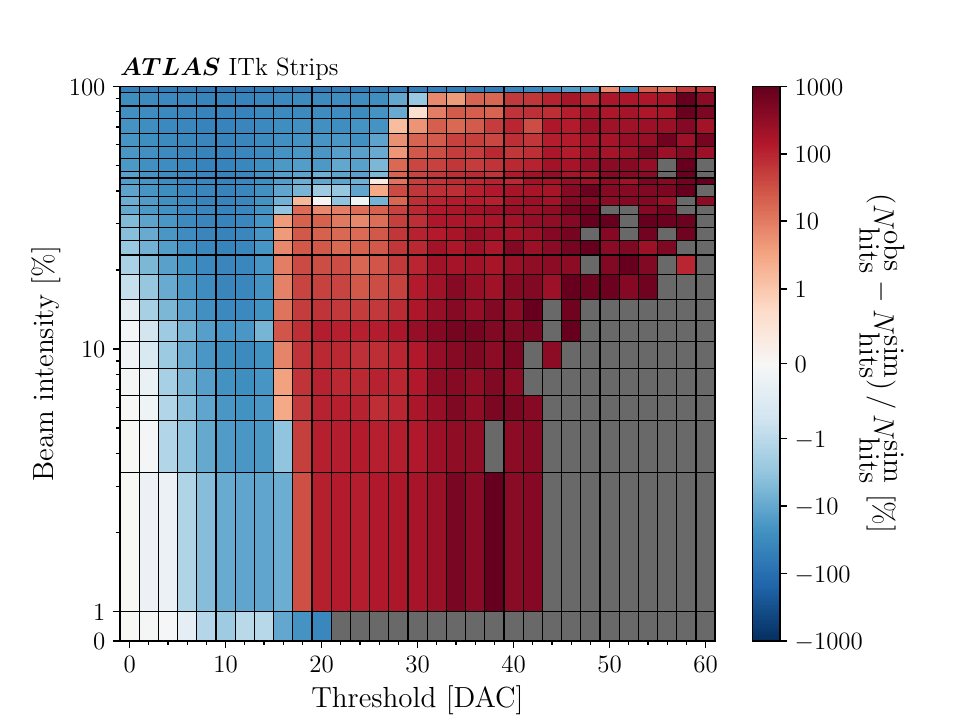}
        \caption{Excluding threshold bounce effects}
    \end{subfigure}
    \caption{The relative difference in the number of observed and simulated hits for all fluxes and thresholds, shown for the simulations (a) including and (b) excluding threshold bounce effects. The grey cells indicate the fluxes/thresholds where the division is undefined due to a lack of simulated hits or where the associated relative difference values go off the scale of the colourbar, also due to a lack of simulated hits. In (b), the colourbar uses a logarithmic scale except in the range of $[-1, 1]\%$, where a linear scale is used instead. In both (a) and (b), the $y$-axis uses a logarithmic scale except in the range of $[0, 1]\%$, where a linear scale is used instead.}
    \label{fig:sim_vs_obs}
\end{figure}

The observed and simulated hits --- both including and excluding threshold bounce effects --- for a number of example beam intensities are shown in figure~\ref{fig:nhits_sim_vs_obs_3x3}. Figure~\ref{fig:sim_vs_obs} shows the relative difference between the observed and simulated hits for all intensities and thresholds. Generally, there is good agreement between data and the MC including threshold bounce effects. The MC consistently exceeds data at low thresholds; however, the difference is small ($<$10\%). Notably, the MC captures the ``smearing'' of the hit profile across a wide range of thresholds. There is a single beam intensity, 11.9\%, where greater data-MC disagreement is observed --- this run is assumed to be an anomaly. In contrast, there is poor agreement between data and the MC excluding threshold bounce effects. This is evident at high beam intensities ($>$10\%), where MC exceeds data at thresholds \unit[$<$15]{DAC}, as well as at low beam intensities, where MC exhibits a deficit at thresholds of \unit[20--30]{DAC}. At the lowest beam intensities, the data and both flavours of simulation are in general agreement with one other. This is expected, as the low-intensity regime contains photons which are spaced far apart. It is clear that accounting for threshold modulation induced by the impulse response of the FE electronics is \emph{crucial} for simulating the hit efficiencies observed.

\begin{figure}[htbp]
    \centering
    \includegraphics[width=\textwidth]{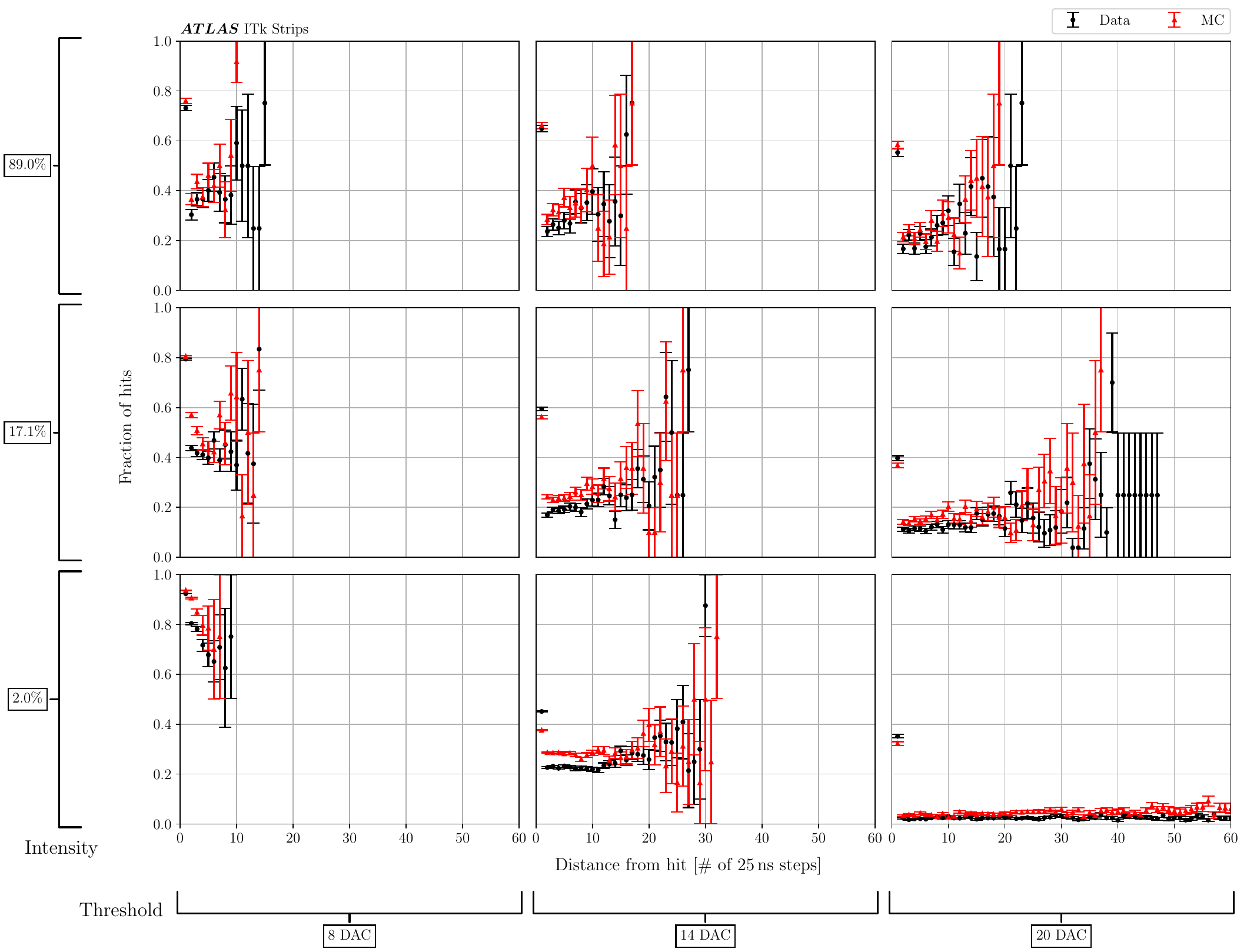}
    \caption{The observed and simulated fractions of triggers which measure a hit as a function of the distance following a hit for nine combinations of beam intensities (89\%, 17.1\%, and 2\%) and discriminator thresholds (8, 14, and \unit[20]{DAC}). The simulation includes threshold bounce effects. The error bars correspond to the Wilson confidence intervals.}
    \label{fig:frac_sim_vs_obs_3x3}
\end{figure}

Figure~\ref{fig:frac_sim_vs_obs_3x3} shows the observed and simulated fractions of triggers which measure a hit as a function of the distance following a hit for the same beam intensities and discriminator thresholds as shown in figure~\ref{fig:fits_3x3}. The simulation includes threshold bounce effects. Adequate data-MC agreement is observed, improving with increasing intensity and threshold. In particular, the MC undershoots or overshoots data in the $\textrm{distance} = 1$ bin and overshoots data in all subsequent bins. However, the simulation does roughly capture the observed slopes, including the regions where the slopes are negative --- for example, at $\textrm{intensity} = 2.0\%$ and $\textrm{threshold} = \unit[12]{DAC}$.

It's worth noting that threshold bounce effects are expected to have a negligible impact on the operation of the ITk detector at the HL-LHC. The ABCStar ASICs are designed for a maximum channel occupancy of 10\%. At maximum channel occupancy, only one in every ten triggers is measuring a hit, so evaluating the impulse response at $t = 9 \times \unit[25]{ns}$ yields $-8\%$ of the maximum response, roughly the same order-of-magnitude as the expected noise. But the occupancy may be even less during operation. Given the worst-case expected particle flux --- including pile-up collisions, \unit[$\mathcal{O}(10^{-3})$]{hadrons/mm$^2$/\unit[25]{ns}} --- and assuming strip pitches and lengths \unit[$\mathcal{O}(0.1)$]{mm} and \unit[$\mathcal{O}(10)$]{mm}, respectively, the average channel occupancy is found to be $\mathcal{O}(0.1\%)$. This calculation additionally assumes each hadron results in a hit. Evaluating the impulse response at $t = 999 \times \unit[25]{ns}$ yields zero; in other words, the current hit will have no impact on the subsequent hit.

\FloatBarrier

\section{Conclusion}
\label{sec:conclusion}

We have presented a measurement of threshold bounce --- occupancy-dependent modulation of the discriminating threshold --- in ATLAS ITk strip modules using a micro-focused photon beam at the B16 Test Beamline at the Diamond Light Source synchrotron. We have shown the impact of threshold bounce to be largest at high photon fluxes and low discriminator thresholds, in line with expectation. In order to confirm that the observed effect can be attributed to the modulation of the readout threshold over several \unit[100]{ns}, we have also presented a Monte Carlo simulation which attributes threshold bounce to the superposition of the impulse response of the module's FE across multiple readout windows. The simulation faithfully reproduces the observed hit efficiencies for varying photon flux and discriminator threshold.

The impact of threshold bounce is significant during beam tests with high local occupancy, and so users must take care to understand if they are in a regime where this effect is present. It was confirmed that threshold bounce is not expected to impact the operation of the ATLAS ITk strip detector at the HL-LHC due to the much lower hit occupancy. However, should unexpected conditions arise at the HL-LHC, the analysis presented in this paper may be used to help understand the effect of threshold bounce on the strip detector's front-end behaviour.

\acknowledgments
\addcontentsline{toc}{section}{Acknowledgements}

This work has been supported by the Canada Foundation for Innovation, the Natural Sciences and Engineering Research Council of Canada, and the Science and Technology Facilities Council (grant number ST/R002592/1).

We thank the Diamond Light Source for access to the B16 Test Beamline (proposal number MM28368), which contributed to the results presented here. The authors would like to thank the personnel of the B16 Test Beamline, especially Oliver Fox and Andy Malandain for providing advice, support, and maintenance during the experiment. We also thank Claudia Gemme and Vitaliy Fadeyev for their review of the manuscript; discussions with Vitaliy were also helpful for properly simulating the operation of the detector.

\appendix
\section{Derivation of the ABCStar FE's impulse response}
\label{app:derivation}

In this appendix, the ABCStar FE's impulse response is derived. The corresponding transfer function is built using the transfer functions for low- and high-pass filters:

\begin{equation}
    \begin{split}
        \textrm{First-order low-pass filter: } & H_\textrm{1LP}(\omega \,;\, \tau) \equiv \frac{1}{1+j\omega\tau} \,, \\
        \textrm{Second-order low-pass filter: } & H_\textrm{2LP}(\omega \,;\, \tau_1,\tau_2) \equiv \frac{1}{1+j\omega\tau_1}\frac{1}{1+j\omega\tau_2} \,, \\
        \textrm{First-order high-pass filter: } & H_\textrm{1HP}(\omega \,;\, \tau) \equiv \frac{j\omega\tau}{1+j\omega\tau} \,. \\
    \end{split}
\end{equation}

\noindent The preamplifier's feedback RC, the preamplifier's bandwidth RC, and the inverting amplifier are each equivalent to a first-order low-pass filter. The ABCStar FE's transfer function is given by the product of the transfer functions for each constituent filter:

\begin{equation}
    \begin{split}
        H_\textrm{ABCStar}(\omega) & = H_\textrm{1LP}(\omega \,;\, \tau_1) \times H_\textrm{1LP}(\omega \,;\, \tau_2) \times H_\textrm{1LP}(\omega \,;\, \tau_3) \\
        & \,\,\,\,\,\, \times H_\textrm{2LP}(\omega \,;\, \tau_4, \tau_5) \times H_\textrm{1HP}(\omega \,;\, \tau_6) \times H_\textrm{1LP}(\omega \,;\, \tau_7)\\
        & = \sum_{i=1}^{7} \frac{C_i}{1+j\omega\tau_i} \,. \\
    \end{split}
\end{equation}

\noindent In the second line, a partial fraction decomposition is used, where $\qty{C_i \,;\, i=1,\ldots,7}$ are constants which depend on $\qty{\tau_i \,;\, i=1,\ldots,7}$. In particular:

\begin{equation}
    \begin{split}
        C_1 & = \frac{\tau_1^5 \times \tau_6}{(\tau_1 - \tau_2) \times (\tau_1 - \tau_3) \times (\tau_1 - \tau_4) \times (\tau_1 - \tau_5) \times (\tau_6 - \tau_1) \times (\tau_1 - \tau_7)} = -1.34 \,, \\
        C_2 & = \frac{\tau_2^5 \times \tau_6}{(\tau_2 - \tau_1) \times (\tau_2 - \tau_3) \times (\tau_2 - \tau_4) \times (\tau_2 - \tau_5) \times (\tau_6 - \tau_2) \times (\tau_2 - \tau_7)} = 26.05 \,, \\
        C_3 & = \frac{\tau_3^5 \times \tau_6}{(\tau_3 - \tau_1) \times (\tau_3 - \tau_2) \times (\tau_3 - \tau_4) \times (\tau_3 - \tau_5) \times (\tau_6 - \tau_3) \times (\tau_3 - \tau_7)} = 0.61 \,, \\
        C_4 & = \frac{\tau_4^5 \times \tau_6}{(\tau_4 - \tau_1) \times (\tau_4 - \tau_2) \times (\tau_4 - \tau_3) \times (\tau_4 - \tau_5) \times (\tau_6 - \tau_4) \times (\tau_4 - \tau_7)} = -0.31 \,, \\
        C_5 & = \frac{\tau_5^5 \times \tau_6}{(\tau_5 - \tau_1) \times (\tau_5 - \tau_2) \times (\tau_5 - \tau_3) \times (\tau_5 - \tau_4) \times (\tau_6 - \tau_5) \times (\tau_5 - \tau_7)} = -26.30 \,, \\
        C_6 & = \frac{-\tau_6^6}{(\tau_6 - \tau_1) \times (\tau_6 - \tau_2) \times (\tau_6 - \tau_3) \times (\tau_6 - \tau_4) \times (\tau_6 - \tau_5) \times (\tau_6 - \tau_7)} = -1.17 \,, \\
        C_7 & = \frac{\tau_6 \times \tau_7^5}{(\tau_7 - \tau_1) \times (\tau_7 - \tau_2) \times (\tau_7 - \tau_3) \times (\tau_7 - \tau_4) \times (\tau_7 - \tau_5) \times (\tau_6 - \tau_7)} = 2.45 \,. \\
    \end{split}
\end{equation}

\noindent Therefore, the ABCStar FE's impulse response, up to an absolute scaling factor, is given by:

\begin{equation}
    V_\textrm{ABCStar}(t) = \mathcal{L}^{-1}\qty{H_\textrm{ABCStar}(\omega)} = \sum_{i=1}^7 \frac{C_i}{\tau_i}\exp\qty(\frac{-t}{\tau_i}) \,,
\end{equation}

\noindent where $\mathcal{L}^{-1}\qty{\ldots}$ denotes the inverse Laplace transform.

\bibliographystyle{JHEP}
\bibliography{bibliography}

\end{document}